\documentclass{PoS}
\usepackage{subfigure, epsfig, rotate, epsf,setspace, amssymb, amsxtra, placeins,slashed}
\newcommand{\msbar}{\overline{\mathrm{MS}}}

\newcommand{\res}{\mathrm{res}}

\title{Continuum Results for Light Hadronic Quantities using Domain Wall Fermions with the Iwasaki and DSDR Gauge Actions}

\ShortTitle{Continuum Results for Light Hadronic Quantities using Domain Wall Fermions with the Iwasaki and DSDR Gauge Actions}

\author{\speaker{Christopher Kelly}
	\\
        Columbia University,\\
        905 Pupin Hall,\\
	116th St \& Broadway,\\
	New York, NY 10027, USA.\\
        E-mail: \email{ckelly@phys.columbia.edu}}

\author{For the RBC and UKQCD collaborations}

\abstract{We present preliminary continuum results for light hadronic quantities obtained by the RBC/UKQCD collaboration using domain wall fermions with both the Iwasaki and the novel Dislocation Suppressing Determinant Ratio (DSDR) gauge actions. The DSDR action allows us to simulate at near physical quark masses on a larger, coarser lattice ($a^{-1} = 1.4$ GeV, $L = 4.6$fm) while retaining good chiral symmetry properties. We discuss our ongoing combined analysis of the three ensemble sets and give early results for the pion and kaon decay constants, quark masses and $B_K$.}

\FullConference{ The XXIX International Symposium on Lattice Field Theory - Lattice 2011\\
July 10-16, 2011\\
Squaw Valley, Lake Tahoe, California}

\begin{document}
\vspace{-1.5cm}
\section{Introduction}
\vspace{-0.3cm}
In these proceedings we present continuum limit predictions for the neutral kaon mixing parameter $B_K$, the strange and average up/down quark masses and the pion decay constant $f_\pi$, obtained via simultaneous chiral/continuum fits to the three domain wall fermion ensemble sets given in table~\ref{tab-ensproperties}. Our most recently generated ensemble set, referred to by the label `32ID', uses a modified Iwasaki gauge action that includes the Dislocation Supressing Determinant Ratio (DSDR) term, which allows us to simulate with near-physical pion masses on a coarser $\beta =1.75$ lattice, while retaining good chiral symmetry and topology tunneling.

\begin{table}[tb]
\vspace{-0.4cm}
\centering
\small{
\begin{tabular}{c|c|c|c|c|c|c|c}
\hline\hline
\rule{0cm}{0.4cm}Label & Size & $S_G$ &  $\beta$ & $N^{\mathrm{ens}}$ & $m_\pi^{\mathrm{uni.}}$ (MeV) & \# configs. & $m_\pi^{PQ} \geq$ (MeV)\\
\hline
\rule{0cm}{0.4cm}32ID & $32^3\times 64\times 32$ & Iwasaki+DSDR 	& 1.75 & 2 & $170$, $250$ 		& 181, 148 & 140\\
\rule{0cm}{0.4cm}24I & $24^3\times 64\times 16$ & Iwasaki 		& 2.13 & 2 & $330$, $420$ 		& 202, 178 & 240\\
\rule{0cm}{0.4cm}32I &$32^3\times 64\times 16$ & Iwasaki 		& 2.25 & 3 & $290$, $350$, $400$ 	& 300, 312, 252 & 220\\
\end{tabular}
}
\caption{A summary of the properties of the three ensemble sets used in this analysis. Here `$S_G$` denotes the gauge action, `$N^{\mathrm{ens}}$' the number of ensembles, `$m_\pi^{\mathrm{uni.}}$' the unitary pion mass on each of those ensembles, `\# configs' the number of gauge configurations used in this analysis, and $m_\pi^{PQ} \geq$ the lightest available partially-quenched pion mass on the ensemble set.}
\label{tab-ensproperties}
\end{table}

Combined analyses of the 24I and 32I lattices, which we refer to collectively as the `2010 analysis', have recently been published~\cite{Aoki:2010dy,Aoki:2010pe}. In these papers we developed a strategy for the combined analysis of multiple ensemble sets that maximises the use of the available data in constraining the fits. In these proceedings we develop this strategy further to include 32ID ensemble set. Note that the number of configurations available on the lightest 32ID ensemble set has almost doubled since the conference, and the heavier ensemble has also increased in size by $35\%$, hence the conclusions presented here differ slightly from those given at the conference.

The layout of these proceedings is as follows: We provide a short discussion on the DSDR term, followed by a summary of our combined fitting strategy. We then present results for the pion decay constant before briefly touching on the non-perturbative renormalisation techniques used in the calculations of $B_K$ and the physical quark masses. Results for these quantities follow. We conclude with a brief summary and outlook for this analysis.
\vspace{-5mm}
\section{The DSDR Term}
\vspace{-2mm}
The 2010 analysis was performed to data over the range $m_\pi = 290 - 420$ MeV. Here the extrapolations down to the physical pion mass of $\sim 135$ MeV provided the dominant contribution to the systematic errors on the continuum predictions. For example, the chiral extrapolation systematic on $f_\pi$ was $\sim 4\%$. This provided a strong motivation for reaching down to lighter quark masses. 

In order to avoid finite-volume effects, simulations with lighter quark masses require lattices with a larger physical volume. However, at our typical couplings the required increase in the number of lattice sites is beyond the reach of our current computing resources; we are therefore forced to simulate with coarser lattices. This has the unfortunate side-effect of increasing the size of the chiral symmetry breaking effects in the domain wall fermion formulation, due to the following mechanism. 

The size of the chiral symmetry breaking is parameterised by an additive mass-shift known as the `residual mass' $m_\res$. This quantity is governed by the eigenvalue density $\rho$ of the four-dimensional Hamiltonian $H_T = 2\tanh^{-1}\left( \frac{H_W}{2+D_W} \right)$
describing quark propagation through the fifth dimension. Here $D_W$ is the Wilson Dirac operator and $H_W = \gamma^5 D_W$ is the hermitian Wilson Dirac operator. The equation above implies a relationship between the eigenmodes of $H_T$ and $H_W$. The modes of the latter are divided into two regions by a mobility edge $\lambda_c$: those above $\lambda_c$ are extended over the lattice whereas those below are localized - in fact this property is essential in ensuring that the four-dimensional effective theory of the fields on the boundary is local. The structure of eigenmodes of $H_W$ implies that $m_\res$ has the following dependence on $L_s$~\cite{Antonio:2008zz}:
\begin{equation}
m_{\mathrm{res}} \sim R_e^4  \rho(\lambda_c)\frac{e^{-\lambda_c L_s}}{L_s} + R_l^4\rho(0)\frac{1}{L_s}\,,
\end{equation}
where the first term contains contributions from the extended modes and the second term from the localized near-zero modes.

\begin{figure}[t]
\vspace{-0.7cm}
\centering 
\includegraphics*[width=0.35\textwidth]{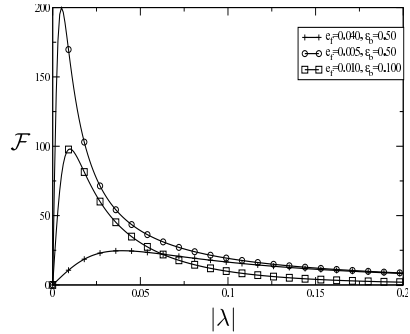}
\caption{An illustration of the molecular dynamics force imparted by the DSDR term at various values of $\epsilon_f$ and $\epsilon_b$, reproduced from ref.~\cite{Renfrew:2009wu} }
\label{fig-dsdrforce}
\end{figure}

In modern simulations, $L_s$ is typically large enough that $m_\res$ is dominated by the near-zero mode contribution. These modes are associated with small tears or `dislocations' in the gauge field, which occur more often as we approach the disordered strong-coupling region. In order to retain good chiral symmetry at stronger coupling we must therefore seek to suppress these modes. However we must retain enough of the very-near-zero modes that allow topological tunneling to occur. This can be achieved by introducing a weighting term, the DSDR term, into the gauge action, given by~\cite{Vranas:1999rz,Vranas:2006zk,Renfrew:2009wu}
\begin{equation}
 \mathcal{W}(M;\epsilon_f;\epsilon_b) = \frac{\mathrm{det}\left[D_W(-M+i\epsilon_b\gamma^5)^\dagger D_W(-M+i\epsilon_b\gamma^5)\right]  }{\mathrm{det}\left[D_W(-M+i\epsilon_f\gamma^5)^\dagger D_W(-M+i\epsilon_f\gamma^5)\right]} = \prod_i \frac{\lambda_i^2+\epsilon_f^2}{\lambda_i^2+\epsilon_b^2}\,,
\end{equation}
where $\lambda_i$ are eigenvalues of $H_W$ and $\epsilon_f$ and $\epsilon_b$ are tunable parameters. This introduces a force in the molecular dynamics evolution of the form
\begin{equation}
\mathcal{F}_i(\epsilon_f,\epsilon_b) = \frac{d}{d\lambda_i}\left(-\log \frac{\lambda_i^2+\epsilon_f^2}{\lambda_i^2+\epsilon_b^2}\right)\,,
\end{equation}
which can be tuned to peak in the near-zero region without further suppressing the very-near-zero modes. In figure~\ref{fig-dsdrforce} we reproduce a plot from ref.~\cite{Renfrew:2009wu} which shows the force as a function of $\lambda$ for several combinations of $\epsilon_f$ and $\epsilon_b$ on a test simulation, demonstrating this suppression.

\vspace{-5mm}
\section{Simultaneous Fitting Procedure}\label{sec-simfits}
\vspace{-2mm}
We obtain our fit forms via a dual expansion to next-to-leading order (NLO) in $a^2$ and the light and heavy quark masses $m_l$ and $m_h$, adopting a power counting that discards terms like $\mathcal{O}(a^2m)$ and $\mathcal{O}(m^2)$ and higher at this order. Expanding around a non-zero mass point $(m_{l0},m_{h0})$ and absorbing constant terms containing $m_{l0}$ into the leading coefficient, we obtain the analytic fit function
\begin{equation}
 f_{ll} = C_0^{f_\pi}\left(1 + C_f a^2\right) + C_1^{f_\pi}\frac{(\tilde m_x + \tilde m_y)}{2} + C_2^{f_\pi}\tilde m_l + C_3^{f_\pi}\left(\tilde m_h - m_{h0}\right)\label{eqn-fpianalytic}
\end{equation}
for the pion decay constant and similar forms for other quantities. Here $\tilde m = m + m_\res$. We also consider expanding about the $SU(2)$ chiral limit (with and without finite-volume corrections), in which case we obtain the usual NLO ChPT fit forms with an extra $a^2$ and $(m_h-m_{h0})$ term. We refer to these as the ChPT and ChPTFV fit forms. Using these three fit ans\"{a}tze, we simultaneously fit the 32I, 24I and 32ID ensemble sets, with the coefficients shared between all three data sets. For ensemble sets $i$ other than the primary set, chosen as the 32I set, we must include appropriate factors of the ratio of lattice spacings $R_a = a^{\rm 32I}/a^i$ and quark masses $R_{l/h} = \tilde m_{l/h}^{\rm 32I}/\tilde m_{l/h}^{\rm i}$ in the fit forms such that the coefficients $C$ can be assumed equal. In this analysis we adopt the so-called `generic scaling' approach~\cite{Aoki:2010dy} in which these ratios are determined as free parameters in the fit. 

Before taking the continuum limit, we must fix the primary lattice spacing $a^{\rm 32I}$ and the quark masses $m_{u/d}^{\rm 32I}$ and $m_s^{\rm 32I}$ in 32I normalisation. This is achieved by varying these parameters until the continuum values of the pion, kaon and Omega baryon masses agree with their physical values. These quantities define the `scaling trajectory' along which the continuum limit is defined. Note that this makes the $a^2$ coefficients of these quantities zero by definition. As the above conditions are applied in the continuum limit, the procedure is necessarily iterative.

The inclusion of the Iwasaki+DSDR data complicates the situation slightly over the 2010 analysis, in that the coefficients of the $a^2$ terms must now be allowed to differ between the two different gauge actions. Although this introduces one extra degree of freedom per quantity into the fits, any algorithmic instability that this may cause is offset by the increased number of data points.

Recall that eqn.~\ref{eqn-fpianalytic} contains a term in the heavy sea-quark mass $m_h$. We vary this parameter using reweighting~\cite{Jung:2010jt,Aoki:2010dy}, whereby the weight of a given gauge configuration is re-evaluated in the path integral at several strange quark masses shifted by up to $20\%$ from the simulated value. This allows us to explore the strange sea quark dependence with minimal cost and to ultimately quote predictions at the physical strange quark mass at the penalty of an increase in statistical error.

\vspace{-5mm}
\section{Preliminary Results For $f_\pi$}
\vspace{-2mm}
$f_\pi$ is obtained from the $\langle 0|A_0|\pi \rangle$ matrix elements in the usual way, the only difference being that for domain wall fermions one must correctly renormalise the 4d axial current to match the continuum current. Following ref.~\cite{Aoki:2010dy}, the renormalisation factor is obtained from the ratio of the 5d DWF conserved \textit{vector} current to the 4d vector current: this was shown to be more precise than the ratio of axial currents due to the unknown renormalisation coefficient between the 5d DWF PCAC current and the continuum current.

\begin{table}
\vspace{-0.4cm}
\small{
\begin{tabular}{c|c|c|c}
\hline\hline
					& 2010 analysis             & This analysis (all data) & This analysis ($m_\pi \leq 350$ MeV)\\
\hline
\rule{0cm}{0.4cm} $f_\pi$ (MeV)  & 124(2)(5)(2) & 125(2)(2)(2) & 127(3)(0.5)(3)\\
\rule{0cm}{0.4cm} $m_{u/d}^{\msbar}(2\ {\rm GeV})$ (MeV) & 3.59(13)(12)(6)(8)        & 3.48(6)(7)(3)(8)         & 3.39(9)(4)(2)(7)     \\
\rule{0cm}{0.4cm} $m_s^{\msbar}(2\ {\rm GeV})$ (MeV) & 96.2(1.5)(0.2)(0.1)(2.1) & 94.9(1.2)(1.4)(0.2)(2.1) & 94.2(1.9)(1.0)(0.4)(2.1)  \\
\rule{0cm}{0.4cm} $\hat B_K$ & 0.749(7)(21)(3)(15) & 0.748(6)(15)(4)(15) & 0.751(11)(8)(4)(14)\\
\end{tabular}}
\caption{Results for $f_\pi$, the average up/down quark mass, the strange quark mass and $B_K$. The first column contains the 2010 analysis result~\cite{Aoki:2010dy,Aoki:2010pe}, the second the result obtained in this analysis by fitting to the full range of available data, and the third by fitting only to data with $m_\pi \leq 350$ MeV. The errors are statistical, chiral, finite-volume and NPR errors (where appropriate) respectively.}
\label{tab-predictions}
\end{table}



\begin{figure}[t]
\vspace{-0.8cm}
\centering
\includegraphics[clip=true,trim=0 0 0 30,width=0.45\textwidth]{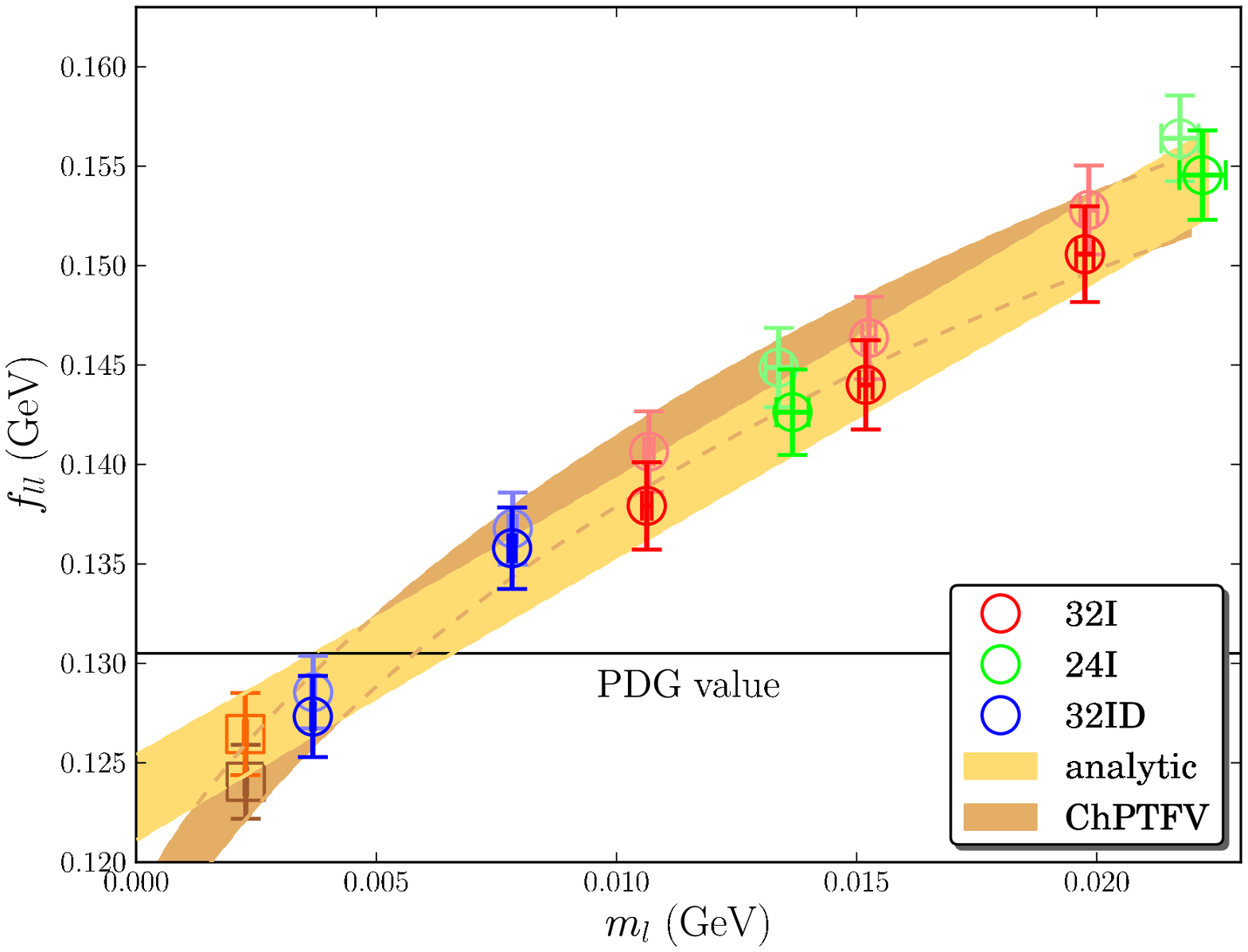}\quad
\includegraphics[clip=true,trim=0 0 0 30,width=0.45\textwidth]{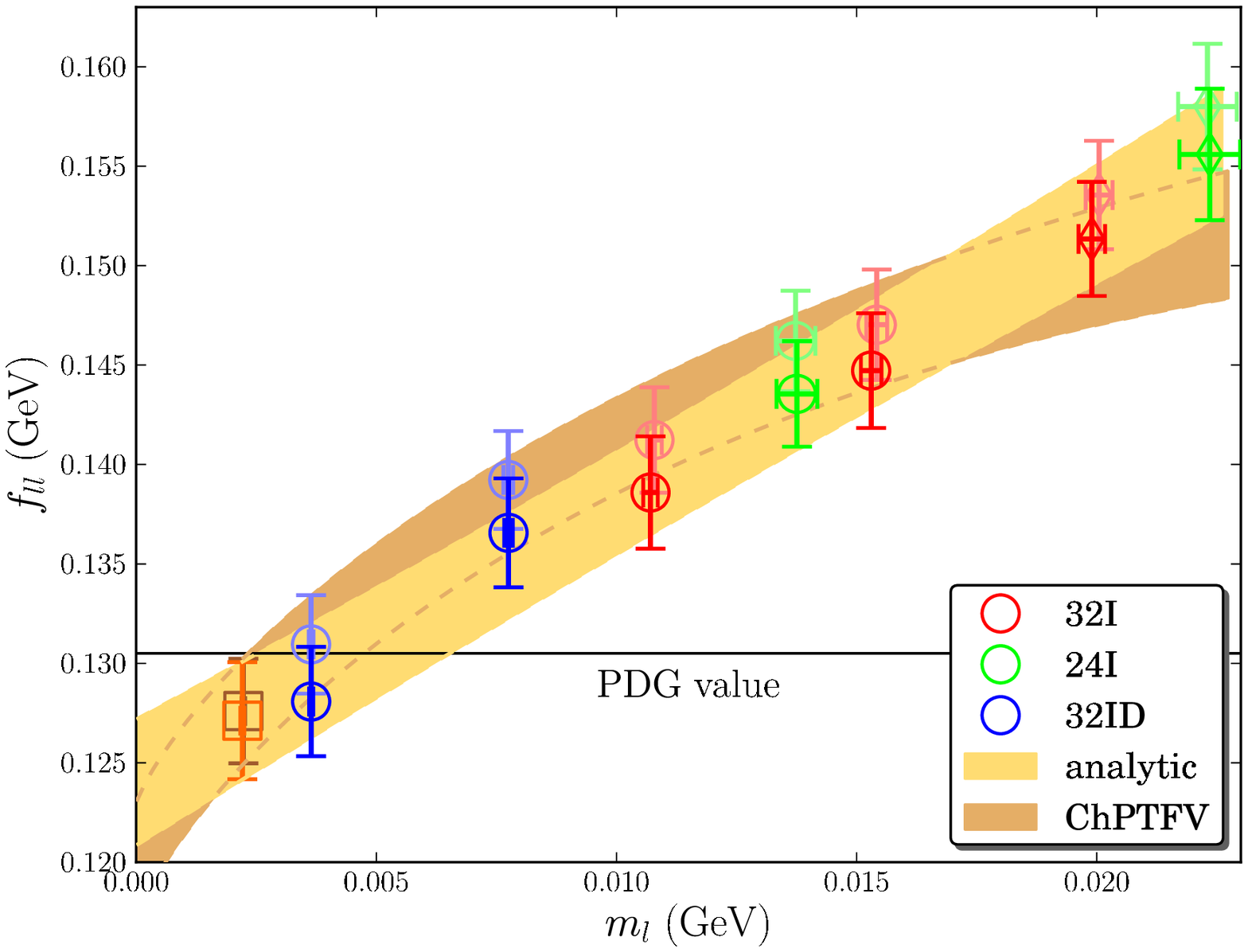}
\caption{Comparisons of the unitary $f_\pi$ data corrected to the continuum limit and fit using the ChPTFV and analytic ans\"{a}tze, where the fits are performed to the full data set (left) and only to the data with $m_\pi\leq 350$ MeV (right). The data included in the fit are marked with circles, and those excluded with diamonds. The continuum limits are marked with squares. Points in pastel colours are corrected to the continuum using the ChPTFV ansatz, and in bold colours by the analytic ansatz.}
\label{fig-fpicomparison}
\end{figure}

Figure~\ref{fig-fpicomparison} shows the chiral extrapolation of $f_\pi$ down to the physical up/down quark mass using the analytic and ChPTFV ans\"{a}tze. Following the 2010 analysis we estimate the error on the chiral extrapolation through the difference of the ChPTFV and analytic predictions, and the finite-volume error from the difference of the ChPTFV and ChPT predictions. We obtain the value given in the second column of table~\ref{tab-predictions}. This result is some $4\%$ ($1.5\sigma$) below the physical value of $130.7$ MeV. A similar discrepancy was noted in the 2010 analysis (first column of the table), and was attributed to the systematic error on the chiral extrapolation. There, the strategy of estimating the chiral error from the difference of the ChPTFV and analytic results produced a systematic error sufficient to explain the discrepancy, but for this analysis it appears to be an underestimate. However, the introduction of the light DSDR data allows us to perform stable fits even after removing some of the heavier data from the 32I and 24I ensemble sets. We can therefore restrict our fits to a region of lighter mass in which we would expect the fit ans\"{a}tze to perform better. Cutting the heaviest two ensembles (cf. table~\ref{tab-ensproperties}) such that the heaviest pion has a mass of $350$ MeV, we obtain the result given in the third column of table~\ref{tab-predictions}. Here, as a result of the cut, the central value has increased such that the result is now consistent with the physical value, even without incorporating the chiral systematic. Notice also that our \textit{ad hoc} chiral systematic has decreased substantially, and the $\chi^2/\mathrm{d.o.f.}$ also decreased, which suggests that the fits do indeed perform better when restricted to this lighter mass range. Although this analysis is still preliminary, these results are promising.

\vspace{-3mm}
\section{Renormalisation for $B_K$ and the quark masses}
\vspace{-2mm}

Before giving results for the quark masses and $B_K$, we present a short summary of the non-perturbative renormalisation techniques that are used to convert these quantities into the canonical $\msbar$ scheme with unprecendented precision.

Direct conversion into the $\msbar$ scheme on the lattice is not possible as this scheme is regularised in non-integer dimensions. Instead we first convert to a convenient intermediate scheme, which is then run to high energy and matched to $\msbar$ using perturbation theory. We use variants of the Rome-Southampton Regularisation-Invariant Momentum (RI-MOM) scheme, in which the renormalisation coefficients are defined from the ratio of the amputated Green's function of an operator to its tree-level value in the Landau gauge at a particular momentum scale. In the RI-MOM scheme, the Green's functions are formed using massless propagators of equal momentum $p$, and the scale is defined as $\mu^2 = p^2$. However this is a so-called `exceptional' momentum configuration in which the hard external momenta can be routed in such a way that parts of the diagram contain only soft momenta - this enhances the effect of the spontaneous chiral symmetry breaking at high momenta. In order to avoid this problem we follow the 2010 analysis in adopting `symmetric' momentum conditions (SMOM), for which the external momenta are not equal but rather obey $\mu^2 = p_1^2 = p^2 = (p_1-p_2)^2$. Note that this applies both for the mass renormalisation, which we obtain from the scalar bilinear vertex, and the four-point $VV+AA$ weak operator contained in the definition of $B_K$; for the latter the Wick contractions comprise traces of two bilinear vertices - the momenta $p_1$ and $p_2$ are assigned to the two propagators forming each of those bilinears. 

As in the 2010 analysis, we use volume source propagators in these calculations, which offer a significant improvement over the traditional point sources, giving statistical errors of the order $0.1\%$ even with only $\mathcal{O}(20)$ gauge configurations.

One of the dominant systematic errors on the NPR in our earlier analyses arose through the use of momenta whose unit vectors were not equivalent under hypercubic rotations, and hence do not have the same discretisation errors; this induces a scatter in the RI-SMOM renormalisation coefficients as a function of momentum. In ref.~\cite{Aoki:2010pe} we described how this can be corrected using twisted boundary conditions in the valence sector, which allow us to smoothly vary the magnitude of the momentum while keeping the direction fixed. We also discussed how another of the dominant systematic errors, that associated with the truncation of the perturbative series, can be reduced by a factor of two by performing the matching to the $\msbar$ scheme at 3 GeV rather than the canonical 2 GeV. For this analysis these improvements have been applied in the case of $B_K$ but have yet to be applied to the quark mass renormalisation.

\vspace{-5mm}
\section{Results for the Quark Masses}
\vspace{-2mm}
The continuum physical quark masses are obtained in the normalisation of the 32I ensemble set, and hence must be renormalised into the $\msbar$ scheme to remove any cutoff dependence. The renormalisation coefficients are determined by performing the continuum extrapolation of $Z_m / R_{l/h}$ over the two Iwasaki lattices, where $R_{l/h}$ are the quark mass ratios defined in section~\ref{sec-simfits} and $R_{l/h}^{\rm 32I}\equiv 1$ by definition. Notice that the coefficient on the 32ID lattice is not needed for this procedure. As mentioned above, the $\msbar(3\ \mathrm{GeV})$ lattice coefficients with twisted boundary conditions have yet to be determined, hence for this analysis we reuse those given in ref.~\cite{Aoki:2010dy}, although the continuum extrapolation is performed anew with the lattice spacings and $R_{l/h}$ obtained here. Following the 2010 analysis procedure, we choose our best NPR scheme for the central value take the truncation error on the renormalisation from the size of the two-loop contribution to the $\msbar$ matching.

From a fit to the full data set, we obtain the values given in the second column of table~\ref{tab-predictions}. For comparison, we give the 2010 analysis result in the first column. We find results that are very consistent, and observe a factor of two reduction in the statistical and chiral systematic errors on the up/down quark mass over the 2010 analysis as a result of including the lighter data. We also see a reduction in the statistical error on the strange quark mass, but also a large \textit{increase} in the chiral error. This is likely a result of allowing the mass ratios $Z_l$ and $Z_h$ to differ between the fit ans\"{a}tze, where before they were fixed to values obtained by matching the lattices at an unphysical mass scale (the fixed trajectory method). We intend to investigate this further.

We also investigate the effect of cutting out the heaviest two ensembles on the quark masses. We obtain the result given in the third column of the table. Here as with $f_\pi$, we see significant improvements in the estimated chiral error, at the expense of an increase in statistical error.


\vspace{-5mm}
\section{Results for $B_K$}
\vspace{-2mm}
We obtain results for $B_K$ through an independent simultaneous fit over our three ensemble sets, using the lattice spacings, quark masses, etc., obtained from the fits above. In particular, we constrain the ChPT fits by including the lowest-order ChPT parameters $f$ and $B$ from the main analysis. The fits are performed to $\msbar$-renormalised data, where we use the coefficients obtained by matching to our lattice scheme at 3 GeV as discussed above, after which we convert to the RGI-scheme for the convenience of the reader. The NPR error is obtained by taking the difference of our two best schemes, as discussed in ref.~\cite{Aoki:2010pe}. From the fits to the full data set, we obtain the result given in the second column of table~\ref{tab-predictions}. This result is very consistent with the 2010 result, and shows a $30\%$ improvement in the chiral error. Fitting only to the data with $m_\pi \leq 350$ MeV, we obtain the result given in the third column of the table. As before we see a substantial decrease in the chiral error, but here the increase in the statistical error is larger than before. This is likely due to the lack of statistical resolution on the lightest 32ID data points.



\begin{figure}[t]
\vspace{-0.8cm}
\centering
\includegraphics[clip=true,trim=0 0 0 30,width=0.45\textwidth]{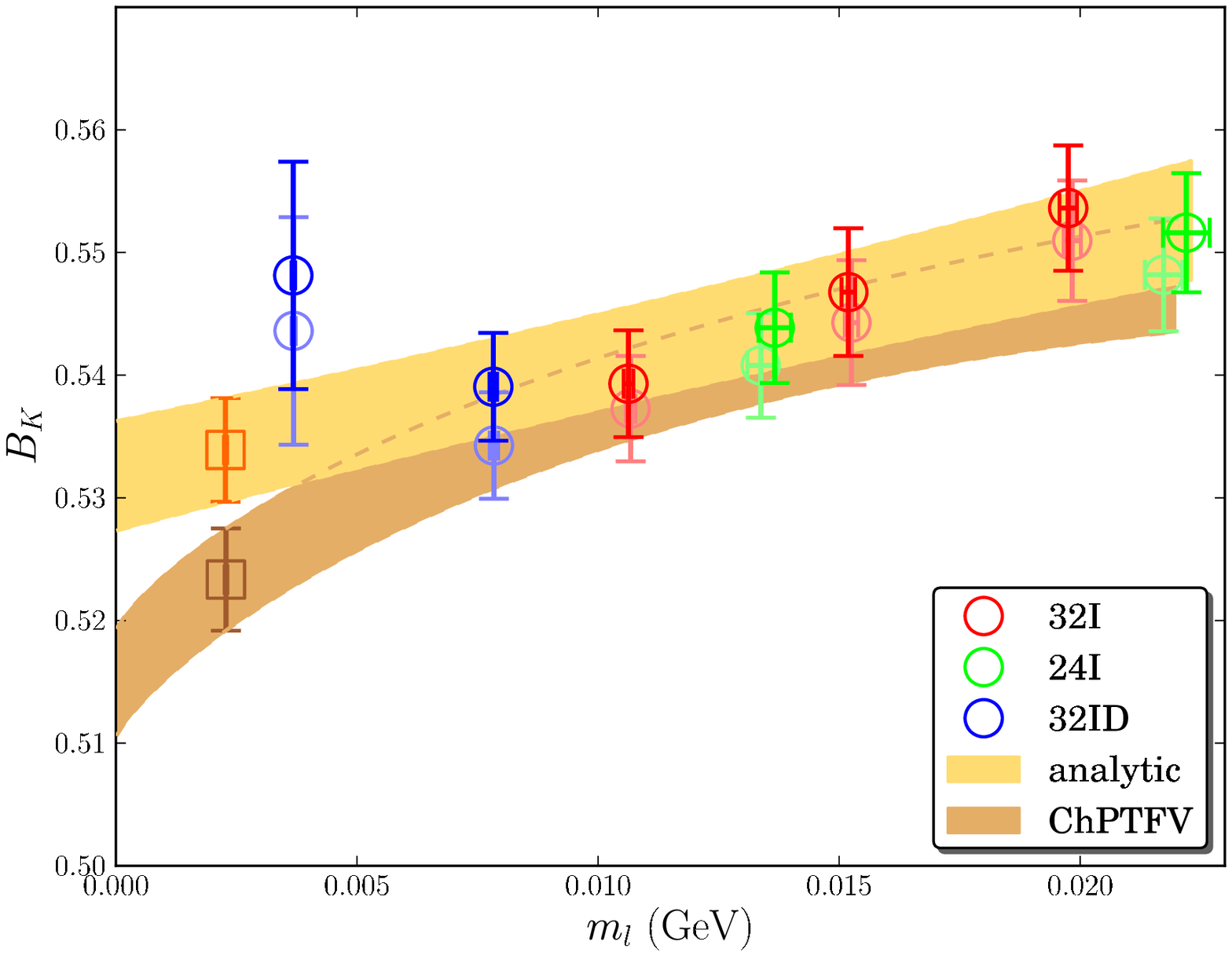}\quad
\includegraphics[clip=true,trim=0 0 0 30,width=0.45\textwidth]{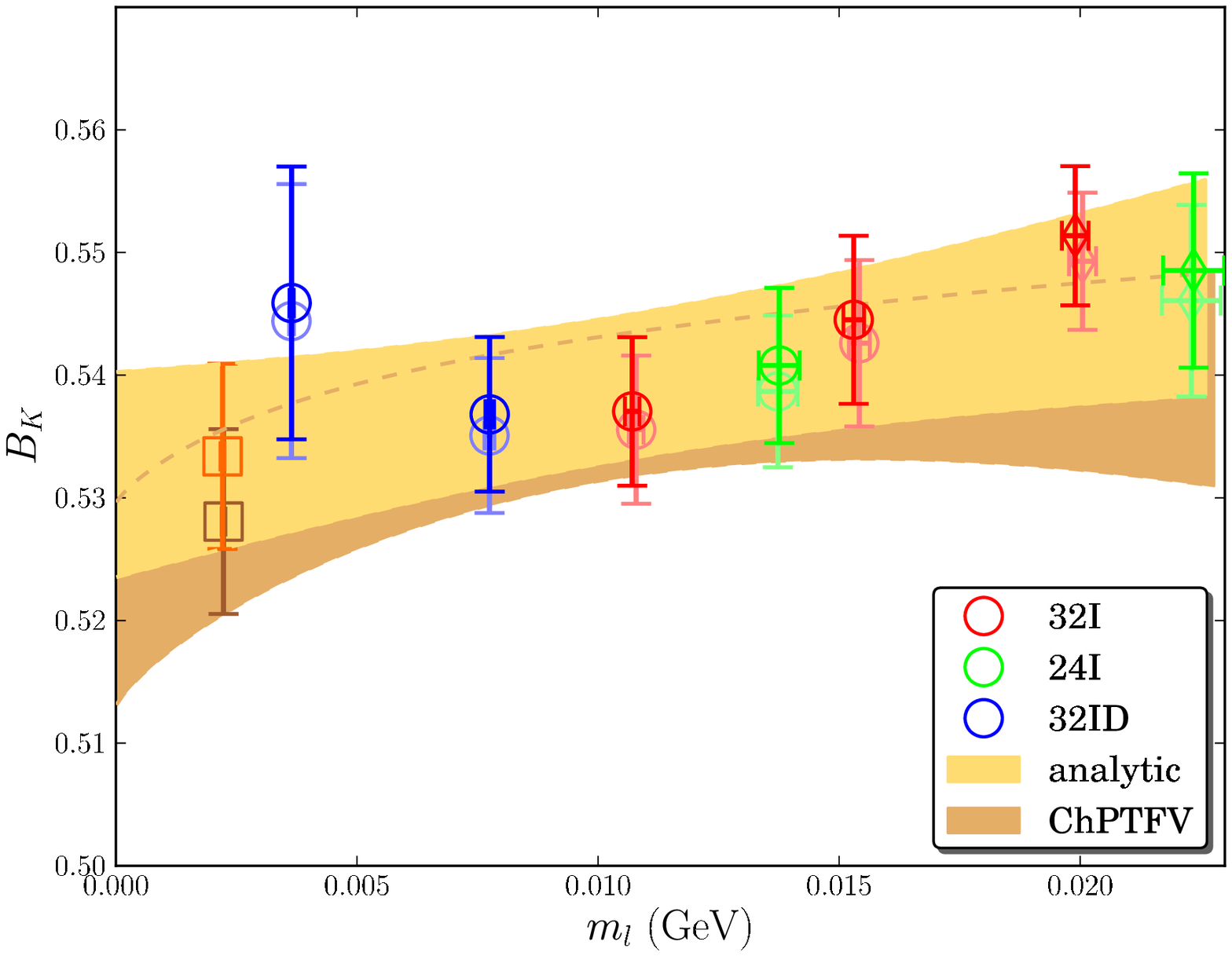}
\caption{Comparisons of the unitary $B_K$ data in the SMOM$(\slashed q,\slashed q)$ scheme corrected to the continuum limit and fit using the ChPTFV and analytic ans\"{a}tze, where the fits are performed to the full data set (left) and only to the data with $m_\pi\leq 350$ MeV (right). The symbols and their colours are described in the caption of figure~\protect\ref{fig-fpicomparison}. }
\label{fig-bkcomparison}
\end{figure}

\vspace{-5mm}
\section{Conclusions and Outlook}
\vspace{-2mm}
Using the Iwasaki+DSDR gauge action, we have been able to simulate with near physical pions while retaining good topological tunneling properties and small finite-volume corrections. Including these data in simultaneous fits with our Iwasaki lattices, we have been able to substantially improve our continuum predictions over the 2010 analysis, especially after cutting out the heaviest two ensembles such that the heaviest pion now has a mass of only $350$ MeV - performing this cut we observed factor of two reductions in our estimated chiral systematic. In addition, our preliminary prediction for $f_\pi$, $127(4)$ MeV, is now consistent with the physical value. We observe that NLO chiral perturbation theory extrapolations over the mass range of $140-350$ MeV should be expected to fail at the 5\% level, hence we expect this value to rise further towards the physical value as we further restrict the fit range in future analyses. 

It is our intention to publish an analysis of these data shortly, after which, alongside continuing to generate more data on the 32ID ensemble set, we intend to commence the generation of further domain wall fermion ensembles with near physical pions, taking advantage of the large increase in computing power provided by our upcoming IBM Blue Gene/Q resources. This will allow us to further refine our continuum predictions.

\end{document}